\documentclass{article}
\usepackage{spconf,amsmath,amssymb,graphicx,hyperref}
\usepackage{booktabs}
\usepackage[table]{xcolor}
\title{segmentwise pruning in audio-language models}
\name{Marcel Gibier, Raphaël Duroselle, Pierre Serrano, Olivier Boeffard, Jean-François Bonastre}
\address{Inria Paris, France}
\begin{document}
%\ninept
%
\maketitle
\begin{abstract}
Recent audio-language models have shown impressive performance across a wide range of audio tasks and are increasingly capable of handling long audio inputs. However, the computing costs in these models heavily depend on sequence length, which can become very large given the nature of audio data. In the vision-language domain, token pruning methods have proven effective in reducing token counts while preserving strong performance on standard benchmarks. In this work, we investigate the relevance and effectiveness of such token selection strategies in the context of audio-language models. We also improve them by proposing a lightweight strategy that takes the time dimension into account. While retaining only a quarter of the initial tokens, our approach results in a relative maximum decrease of 2\% in CIDEr on Clotho v2 and a relative maximum decrease of 4\% in accuracy on MMAU.
% JF : que veut dire "operations in these models heavily..." ? Préciser le "operations"
% Est-ce "operational costs" ?
% PS2 : un mot sur le temps d'inférence? "while resulting in XX % speed up"

\end{abstract}
\section{Introduction}
\label{sec:intro}

\begin{figure}[t]
    \centering
    \includegraphics[width=\columnwidth]{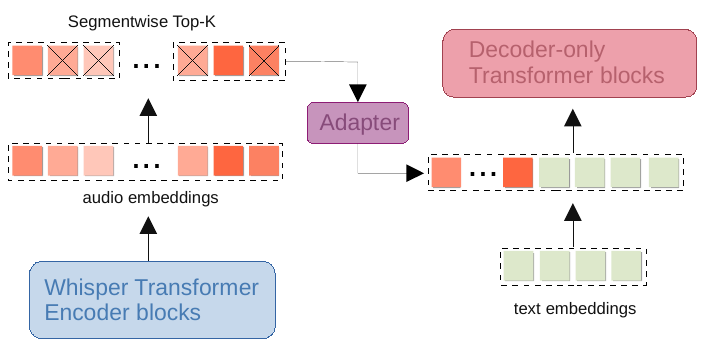}
    \caption{Audio token embeddings are color-coded by aggregate incoming attention. Segmentwise Top-K, with $S=\frac{N}{3}$ and $\lfloor K/S \rfloor=1$ ($S$ = number of segments, $K$ = total tokens retained), selects one token every three tokens from an audio sequence of length $N$.}
    \label{fig:method}
\end{figure}

Recently, audio-language models (ALMs) have demonstrated their ability to perform well across a wide range of tasks, not only in audio understanding, but also in reasoning \cite{mmau}. Although some models aim to reduce the number of computational operations by incorporating cross-attention mechanisms \cite{af2}, the most effective approaches generally rely on a prefix strategy \cite{qwenomni, kimiaudio, af3}. The adopted strategy consists in concatenating the audio representation with the textual embeddings, the projections of the tokenized prompts, prior to processing by the Transformer layers (Figure~\ref{fig:method}). However, the computational complexity of the attention mechanism grows quadratically with the input sequence length. Since the temporal dimension of the audio representation is typically much larger than that of the text \cite{rightproblem, pslm}, this concatenation introduces a substantial computational overhead. 

% PS2 : le "computational operations" pour parler du fonctionnement du LLM peut porter à confusion? Je le retirerais : 
% State-of-the-art model architectures generally rely on a prefix strategy \cite{qwenomni, kimiaudio, af3}. In this approach, the audio representation is concatenated with the textual embeddings (the projections of the tokenized prompts) before being processed by the Transformer layers.
% Marcel : parler de la figure ici like in figure 1

To address this issue, several categories of token compression methods have been developed, which can be broadly grouped into four main families \cite{tokenstalk}:  
\begin{itemize}
    \item \textbf{Transformation-based:} methods that transform tokens into a more compact representation, e.g., using convolution, pooling, or other dimensionality-reduction techniques.  
    \item \textbf{Similarity-based:} methods that merge tokens exhibiting high similarity \cite{tome, visionzip}.  
    \item \textbf{Attention-based:} pruning approaches that discard tokens receiving the least attention \cite{fastv, sparsevlm}.  
    \item \textbf{Query-based}: methods that employ learnable queries to guide the token compression process, i.e., a small set of trainable vectors that attend to the input tokens and extract the most relevant information into a compact representation. \cite{blip2, gama}.  
\end{itemize}

Notably, recent study have demonstrated that token merging can be effectively applied to Audio Spectrogram Transformers, enabling significant compression of audio representations with only a marginal impact on classification performance \cite{fastast}.  

However while most pruning approaches have been developed in the context of vision models, research on audio models remains relatively scarce. Moreover, existing methods rely exclusively on attention scores \cite{audiopruning, speechprune}.

Our main contribution is to demonstrate that pruning is an effective and broadly applicable strategy for ALMs. To this end, we compare pruning methods on two audio understanding tasks, automatic audio captioning and audio question answering, using two state-of-the-art models Whisper-based ALMs, Qwen2-Audio-7B-Instruct \cite{qwen2audio} and Audio Flamingo 3 \cite{af3}. We further introduce an effective pruning method that better accounts for the temporal nature of audio signals. Our approach achieves performance that surpasses existing methods on the Clotho-v2\cite{clotho}, AudioCaps\cite{audiocaps}, ClothoAQA\cite{clothoaqa} and MMAU\cite{mmau} benchmarks.

% PS2 : ta figure 1 n'est pas commentée et tu ne t'en sers plus après dans le texte. Tu pourrais t'en servir dans l'intro ou quand tu détailles ta méthode pour que tout le monde comprenne où tu vas te placer. 

\section{Related work}
\label{sec:related}

% PS2 : modifier la tournure pour ne pas commencer par la citation. 

Pruning in audio understanding tasks using transformer models operating on spectrograms has been investigated by \cite{audiopruning}, specifically with the Audio Spectrogram Transformer (AST) \cite{ast}. These models, inherited from vision architectures such as the Vision Transformer (ViT) \cite{vit}, adopt a structure in which the input is divided into fixed-size patches. Consequently, the number of input tokens is constant, and the number of retained tokens depends solely on the pruning rate applied. Moreover, as in ViTs used for image classification, a special [CLS] token is included and typically serves as a reference point for attention weights.

The study shows that applying the Top-K pruning strategy during model training by retaining only a proportion of the most “dominant” tokens, i.e., those receiving the highest attention from the [CLS] token across the different ViT blocks, reduces the number of tokens by more than half while keeping the accuracy loss below 1\% on a classification benchmark such as ESC-50 \cite{esc}.

In contrast, \textit{SpeechPrune} %RD the second study
\cite{speechprune} does not examine attention within the audio encoder but rather directly in the LLM’s decoder. \textit{SpeechPrune} operates inside the LLM backbone, where it prunes audio tokens based on the attention scores from the very first transformer layer. By leveraging these initial attention signals, the method improves accuracy by nearly 30\% on the benchmark specifically developed for it, while simultaneously  pruning 20\% of the tokens.

In vision, effective methods go beyond simply pruning non-dominant tokens. For example, \textit{VisionZip} \cite{visionzip} preserves dominant tokens, completes coverage by sampling from the rest, and merges redundant tokens using key similarity. While this strategy can be applied at inference time, it generally yields greater performance improvements when coupled with training. 
% jf : Vous etes sût dze vouloir la phrase suivante ?
%Such methods thus leverage both the relevance and redundancy of tokens, leading to more compact and informative representations.
% Si oui, peut-être reformuler :
%These methods offer more compact and informative representations by acting on both the relevance (to be maximized) and redundancy (to be minimized) of selected tokens.
% PS2 : ok pour la retirer 

% PS2 : j'ai l'impression qu'il manque une transition avec la partie Method
% JE pense qu'il ne faut pas appeler cette partie méthode. Mais peut être Segmentwise Top-K Token Pruning directement ?

\section{Method}
\label{sec:method}

We propose a pruning method that requires no training and is applied exclusively at inference time, based on an analysis of the attention in the audio encoder output of ALMs.

%A method that does not require any training and is applied solely at inference time is preferred, which rules out Transformation-based and Query-based approaches.

% PS2 : 
% We propose a pruning method that requires no training and is applied exclusively at inference time, based on an analysis of the attention in the audio encoder output of ALMs.

\subsection{Attention Concentration on Few Tokens}

To motivate the selection of only a subset of tokens, it is important to analyze how attention is distributed over the audio encoder outputs. We refer to these outputs as tokens, although they correspond to the sequence of audio embeddings, in analogy with Vision Transformers that also denote patch embeddings as tokens.

% PS2 : supprimer "Slight abuse"
% We refer to these outputs as tokens, although they correspond to the sequence of audio embeddings, in analogy with Vision Transformers that also denote patch embeddings as tokens.

\begin{figure}[t]
    \centering
    \includegraphics[width=\columnwidth]{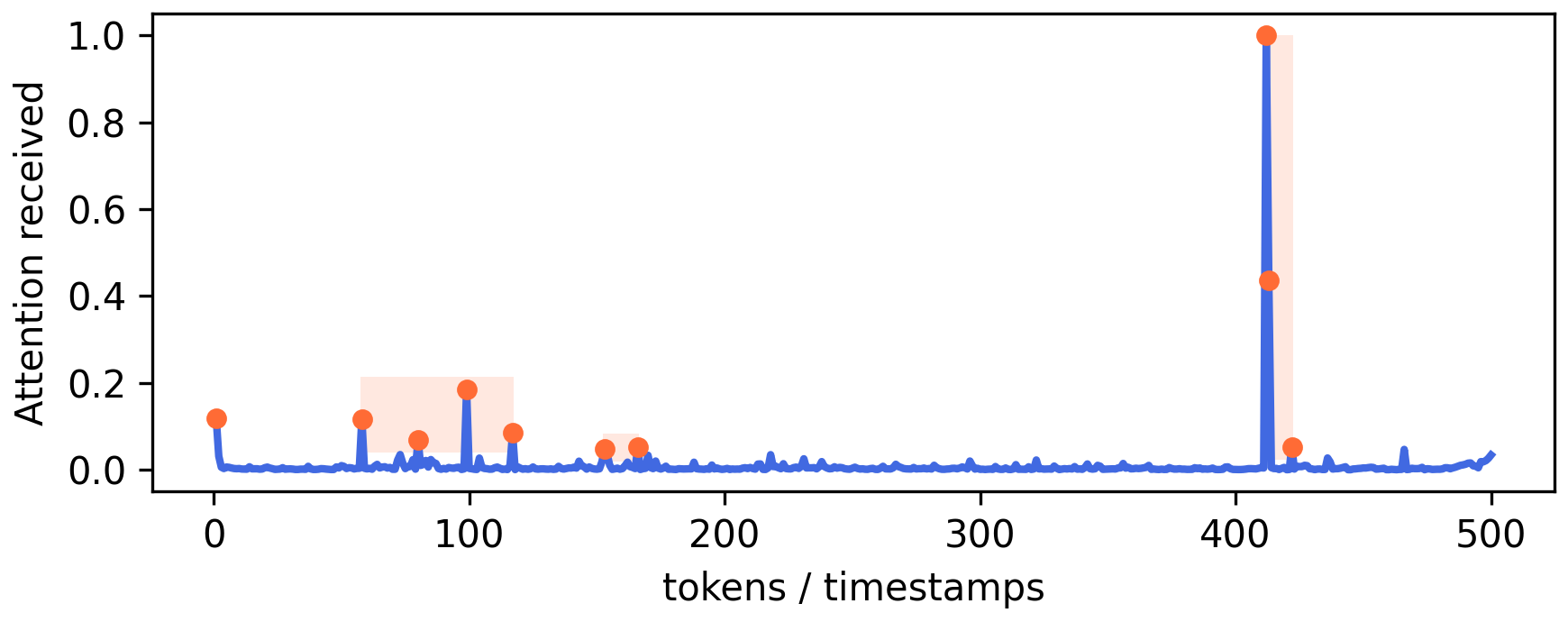}
    \caption{Max-normalized attention received by each token for a 10-second Clotho v2 audio sample. In orange: the 10 most important tokens. Rectangles indicate groups of important tokens that are very close to each other.}
    \label{fig:attention}
\end{figure}

The ALMs under study employ an adapter, typically implemented as a fully connected layer, that projects the audio embeddings into the same dimensional space as the textual embeddings provided to the language model. 
% PS2 : il manque des éléments dans la phrase ci-dessous
Before being passed to the adapter, the audio embeddings are visualized through their attention weights. Figure~\ref{fig:attention} illustrates the attention assigned by the last layer of the Whisper encoder in Audio Flamingo 3 where each token (timestamps) receives attention from the other tokens in the sequence.
% PS2 : Fig. 2 illustrates the attention assigned by the last layer of the Whisper encoder in Qwen2-Audio to each token of the audio embeddings for a given example.

We observe that a small minority of tokens capture most of the attention, likely concentrating the most relevant audio information.
% PS2 : il manque un élément pour accentuer ton point. 
% "Si je sélectionne avec ce seul critère comme avec XX méthode, on perd le reste de l'info. Or elle sera peut être nécessaire pour le LLM suivant la question qui lui est posée."
However, the most important tokens are often close to each other, suggesting limited diversity in the selection method. This motivates mitigating attention scores with another criteria to promote diversity of the audio tokens.

\subsection{Comparison Methods}
To evaluate the effectiveness of pruning in ALMs, we adapt two token-reduction strategies to the audio modality: (i) Top-K selection, a strong attention-based baseline \cite{which}, and (ii) \textit{VisionZip}, an hybrid attention and similarity-based method which achieves state-of-the-art performance among pruning methods for vision–language models. %We apply both as attention-guided pruning in the audio encoder, based on the hypothesis that the downstream LLM consumes already-summarized embeddings and thus lacks sufficient temporal–spectral structure to decide what to discard without incurring information loss.

Top-K, consists in selecting the $K$ tokens in a sequence of length $N$ with the highest attention scores, where attention score of each head $S_h \in \mathbb{R}^{\text{batch size} \times N \times N}$ is defined as:
\[
S_h = \text{Softmax}\!\left(\frac{Q_h K_h^{\top}}{\sqrt{D_h}}\right),
\]
where $D_h$ is the head dimension, and $Q_h$ and $K_h$ represent query and key, respectively.
To obtain a value for each token in the absence of the [CLS] token, the attention scores are summed over the second dimension (attending tokens) and averaged across the heads, which we do for simplicity and efficiency, although other aggregation methods exist \cite{adaptsparse, zerotprune}.

\textit{VisionZip} retains dominant tokens (globally salient) like in Top-K and completes it with contextual tokens obtained via similarity-based aggregation of the remaining tokens in order to summarize residual local details.

\subsection{Segmentwise Top-K Token Pruning }
% PS : l'attention à l'échelle du modèle audio n'est pas la même qu'au niveau du LLM qui va lui aller chercher la bonne information dans son vecteur
% 
The inherently sequential nature of audio signals suggests adopting a pruning strategy that explicitly accounts for their temporal dimension. Furthermore, we hypothesize that tokens receiving the highest attention values do not necessarily correspond to the most relevant information, as multiple tokens may be redundant and encode similar content. To mitigate this redundancy while maximizing informational coverage, we propose segmenting the encoder output into \(S\) non-overlapping temporal intervals and selecting, within each segment, the subset of \(\lfloor K/S \rfloor\) tokens with the highest attention scores with respect to the entire audio sequence, where \(K\) denotes the total number of tokens to retain. Thus, when \(S=1\), we recover the standard Top-K method.

This temporal distribution constraint ensures greater diversity among the retained tokens and promotes a more balanced representation of the audio signal.

\begin{table}[t]
\centering
\footnotesize
\setlength{\tabcolsep}{4pt}
\renewcommand{\arraystretch}{1.2}
\label{tab:af3_retention}
\resizebox{\columnwidth}{!}{%
\begin{tabular}{l|c|c|cc|cccc}
\hline
\textbf{Method} &
\textbf{Clotho-v2} &
\textbf{AudioCaps} &
\multicolumn{2}{c|}{\textbf{ClothoAQA}} &
\multicolumn{4}{c}{\textbf{MMAU}} \\
& & & \textit{unanimous} & \textit{non-binary} & \textit{sound} & \textit{speech} & \textit{music} & \textit{total} \\
\hline

\rowcolor{gray!15}
\multicolumn{9}{c}{\textbf{Retain 100\% Tokens}} \\
vanilla AF3 & 0.50 & 0.67 & 0.91 & 0.50 & 0.80 & 0.66 & 0.74 & 0.73 \\
\hline

\rowcolor{gray!15}
\multicolumn{9}{c}{\textbf{Retain $\downarrow$ \textcolor{green!60!black}{50\,\%} Tokens}} \\
Top-K & 0.48 & 0.65 & 0.89 & 0.49 & 0.78 & 0.57 & 0.73 & 0.69 \\
VisionZip & 0.48 & 0.65 & \textbf{0.90} & \underline{0.50} & 0.77 & 0.56 & 0.73 & 0.69 \\
Segmentwise Top-K & \textbf{0.49} & \textbf{0.66} & \textbf{0.90} & \underline{\textbf{0.59}} & 0.78 & 0.65 & 0.74 & \underline{\textbf{0.73}} \\
\hline

\rowcolor{gray!15}
\multicolumn{9}{c}{\textbf{Retain $\downarrow$ \textcolor{green!60!black}{25\,\%} Tokens}} \\
Top-K & 0.48 & 0.65 & 0.89 & 0.49 & 0.78 & 0.52 & 0.74 & 0.68 \\
VisionZip & 0.48 & 0.65 & 0.89 & 0.48 & 0.77 & 0.50 & 0.73 & 0.67 \\
Segmentwise Top-K & \textbf{0.49} & \textbf{0.66} & \textbf{0.90} & \underline{\textbf{0.52}} & 0.78 & 0.57 & 0.74 & \textbf{0.70} \\
\hline

\rowcolor{gray!15}
\multicolumn{9}{c}{\textbf{Retain $\downarrow$ \textcolor{green!60!black}{10\,\%} Tokens}} \\
Top-K & 0.42 & 0.54 & 0.86 & 0.45 & 0.74 & 0.46 & 0.71 & 0.64 \\
VisionZip & 0.41 & 0.53 & 0.85 & 0.43 & 0.76 & 0.47 & 0.72 & 0.64 \\
Segmentwise Top-K & \textbf{0.45} & \textbf{0.55} & \textbf{0.87} & \underline{\textbf{0.50}} & 0.77 & 0.50 & 0.73 & \textbf{0.67} \\
\hline
\end{tabular}%
}
\caption{Audio Flamingo 3 (AF3) results with progressive token retention. The results on Clotho v2 and AudioCaps are evaluated using the CIDEr metric \cite{cider}, whereas the performance on ClothoAQA and MMAU is assessed in terms of accuracy. Bold indicates the best results for each token retention rate; underlining marks results that are equal to or better than the vanilla model.}
\label{tab:af3}
\end{table}

\begin{table}[t]
\centering
\footnotesize
\setlength{\tabcolsep}{4pt}
\renewcommand{\arraystretch}{1.2}
\label{tab:af3_retention}
\resizebox{\columnwidth}{!}{%
\begin{tabular}{l|c|c|cc|cccc}
\hline
\textbf{Method} &
\textbf{Clotho-v2} &
\textbf{AudioCaps} &
\multicolumn{2}{c|}{\textbf{ClothoAQA}} &
\multicolumn{4}{c}{\textbf{MMAU}} \\
& & & \textit{unanimous} & \textit{non-binary} & \textit{sound} & \textit{speech} & \textit{music} & \textit{total} \\
\hline

\rowcolor{gray!15}
\multicolumn{9}{c}{\textbf{Retain 100\% Tokens}} \\
vanilla Q2A & 0.29 & 0.39 & 0.77 & 0.53 & 0.63 & 0.52 & 0.59 & 0.58 \\
\hline

\rowcolor{gray!15}
\multicolumn{9}{c}{\textbf{Retain $\downarrow$ \textcolor{green!60!black}{50\,\%} Tokens}} \\
Top-K & \underline{\textbf{0.34}} & \underline{0.43} & \underline{0.80} &  \underline{\textbf{0.53}} & 0.60 & 0.48 & 0.58 & 0.55 \\
VisionZip & \underline{\textbf{0.34}} & \underline{\textbf{0.44}} &  \underline{0.80} & 0.51 & 0.63 & 0.48 & 0.58 & 0.57 \\
Segmentwise Top-K & \underline{\textbf{0.34}} & \underline{\textbf{0.44}} &  \underline{\textbf{0.81}} &  \underline{\textbf{0.53}} & 0.61 & 0.51 & 0.61 & \underline{\textbf{0.58}} \\
\hline

\rowcolor{gray!15}
\multicolumn{9}{c}{\textbf{Retain $\downarrow$ \textcolor{green!60!black}{25\,\%} Tokens}} \\
Top-K & \underline{0.32} & \underline{0.46} & \underline{0.78} & 0.52 & 0.56 & 0.46 & 0.56 & 0.53 \\
VisionZip & \underline{0.32} & \underline{0.44} & \underline{\textbf{0.79}} & 0.51 & 0.61 & 0.44 & 0.57 & 0.54 \\
Segmentwise Top-K & \underline{\textbf{0.33}} & \underline{\textbf{0.48}} & \underline{\textbf{0.79}} &  \underline{\textbf{0.53}} &  0.60 & 0.46 & 0.58 & \textbf{0.55} \\
\hline

\rowcolor{gray!15}
\multicolumn{9}{c}{\textbf{Retain $\downarrow$ \textcolor{green!60!black}{10\,\%} Tokens}} \\
Top-K & 0.25 & \underline{0.39} & 0.71 & 0.48 & 0.54 & 0.40 & 0.48 & 0.48 \\
VisionZip & 0.26 & \underline{0.39} & 0.71 & 0.46 & 0.53 & 0.40 & 0.49 & 0.47 \\
Segmentwise Top-K & \textbf{0.27} & \underline{\textbf{0.41}} & \textbf{0.73} & \textbf{0.49} & 0.56 & 0.42 & 0.48 & \textbf{0.49} \\
\hline
\end{tabular}%
}
\caption{Qwen2-Audio-7B-Instruct (Q2A) results with progressive token retention. The caption is the same as in Table~\ref{tab:af3}}
\label{tab:qw2}
\end{table}

% PS2 : préciser que Table 2 même légende que table 1

\section{Experiments and results}
\label{sec:experiments}
\subsection{Models}
In this section, we select two ALMs to assess the relevance of our pruning method and to compare against others: Qwen2-Audio-7B-Instruct and Audio Flamingo 3. The former, as it has been widely adopted by the community \cite{speechprune, audioreasoner}, and the latter, as it achieves state-of-the-art performance across about twenty audio benchmarks. Both use Whisper-large-v3 \cite{whisper} as their audio encoder, which operates on 16 kHz mono waveforms. The raw signal is first converted into a 128-channel log-Mel spectrogram using a 25 ms analysis window and a 10 ms hop size. Consecutive spectrogram frames are then grouped into non-overlapping patches of two frames. This processing yields a sequence of features with a temporal resolution of 50 Hz, each represented by a 1280-dimensional embedding (obtained by linearly projecting). 
The representation is first enriched with positional encodings to retain temporal ordering. Then, the decoder processes the representation using a series of Transformer blocks followed by a pooling operation with stride two. Consequently, a 30-second audio segment (the non-overlapping chunk size used during inference) is mapped to an embedding $\in \mathbb{R}^{750 \times 1280}$.

\subsection{Tasks and Benchmarks}
To assess the relevance of our approach, we evaluate it on two complementary tasks: 
\emph{audio captioning}, which requires the model to summarize the most salient information in the input audio, 
and \emph{audio question answering (AQA)}, which probes more targeted aspects of audio understanding. 
We consider four established benchmarks, two for each task:
\textbf{MMAU (test-mini)} comprising 1,000 audio clips with reasoning-oriented questions and multiple-choice answers, spanning three categories of audio (\emph{sound}, \emph{music}, and \emph{speech}). \textbf{ClothoAQA (test)} consisting of 1,109 audio–question pairs with binary (\emph{yes/no}) answers and an additional 932 pairs with open-ended answers. \textbf{Clotho v2 (evaluation)} containing 1,045 audio–caption pairs. \textbf{AudioCaps (test)} providing 4,876 audio–caption pairs.

% PS2 : ici tu dis avec quoi tu compares Vision Zip, Top-K standard etc

\subsection{Implementation of experiments}

% PS2 : rappeler de quoi tu parles. "In the segmentwise Top-K, we chose...
% retrier "Although the approach...
In the Segmentwise Top-K method, we chose a number of segments $S=10$ to accommodate varying audio lengths; since most clips are 10–30~s long, this yields ~1–3 s windows that are well-suited to capture local information. Although the approach could also have been applied using a fixed window size. We also evaluated performance on MMAU by varying the number of segments between 2 and 15, and found that using 10 segments yields the highest accuracy. A more thorough hyperparameter search on the other benchmarks would likely lead to improved results.

The sequence of tokens obtained by Top-K was constructed in descending order of the attention they received. We also experimented on MMAU by replacing the tokens in temporal order but obtained the same scores, which may be explained by the positional encodings already present in the audio tokens.

In our implementation of \textit{VisionZip}, we adopted the same ratio of contextual to dominant tokens ($\approx 0.18$) as in the original paper.

To ensure easier reproducibility, all outputs were generated via greedy decoding with the models’ original task-specific prompts.

Since it is challenging to directly extract the formatted answer from Qwen2-Audio-7B-Instruct on the MMAU benchmark, we instead rely on the maximum similarity score between the generated response and the candidate options, computed from Sentence-BERT sentence embeddings \cite{sentencebert}.

\subsection{Results}

Performances of the pruning methods applied to Audio Flamingo 3 and Qwen2-Audio-7B-Instruct are reported in Tables~\ref{tab:af3} and \ref{tab:qw2}.
Across both models and all four benchmarks, retaining up to 25\% of the tokens preserves performance: evaluation metrics remain close to the full-token baseline. %(Table~\ref{tab:af3})
In contrast, retaining only 10\% of the tokens generally causes a stronger performance drop. However, this is not always the case. For example, on ClothoAQA we achieve the same score with AF3 even when keeping only 10\% of the tokens. 

Interestingly, we sometimes observe gains, suggesting that pruning removes redundant content and yields a cleaner signal for the LLM’s audio processing. For example, Qwen2-Audio-7B-Instruct achieves up to a 23\% relative improvement in CIDEr score on AudioCaps and a 12\% relative improvement on Clotho v2 compared to the vanilla model.
% PS2 : quel gain ? sur quel corpus ?  sur quelle tache ta méthode est plus efficace ou equivalente aux autres ? 
% PS2 : je pense que tu peux aller plus loin. Par exemple sur le speech dans MMAU tu gagnes beaucoup plus que les autres méthodes et tu n'en parles pas. Pareil dans l'intro tu quantifies des choses et tu n'en parles pas là

% PS2 : retirer modestly
Overall, our method matches or outperforms the two alternatives, with consistent gains across benchmarks and retention rates.

\section{Further Analysis}
\subsection{The importance of attention in the model’s response}

% PS2 : pas Baselines, trouver un autre terme. 
% retirer question-answer leakage qui peut porter à confusion. Tu peux juste dire que tu fais sur le captioning. Tu n'introduis pas CIDEr avant. 

We probe attention’s role by comparing our method to two illustrative variants: Random (same token budget, sampled across frames) and Bottom-K (least-attended tokens; Table~\ref{tab:random}). We evaluate only captioning. With Random, CIDEr drops by 14\% and 16\% when keeping 25\% and 10\% of tokens, showing small but consistent gains of our method over Random. With Bottom-K, CIDEr falls by 90\% and 95\%, confirming least-attended tokens are insufficient. As compression eases, Random converges toward Top-K as informative tokens are more likely to be sampled.

% PS2 : La je perds le fil tu cites Sorted Top-K etc et pas segmentwise ? vérifier le texte et les légendes 

\begin{table}[h!]
\centering
\resizebox{0.6\columnwidth}{!}{%
\begin{tabular}{lcccc}
\toprule
\textbf{Method} & \textbf{Ours} & \textbf{Random} & \textbf{Bottom-K} \\
\midrule
\textbf{50\%} & 0.49 & 0.46 & 0.12 \\
\textbf{25\%} & 0.49 & 0.42 & 0.05 \\
\textbf{10\%} & 0.45 & 0.37 & 0.02 \\

\bottomrule
\end{tabular}
}
\caption{AF3 results for the Random and Bottom-K methods on Clotho v2 keeping only 10\% and 25\% of the audio tokens.}
\label{tab:random}
\end{table}

\subsection{Efficiency Analysis}
To confirm that there is indeed a speed gain during inference, we report computation time during the prefill phase and the decoding phase (Table~\ref{tab:time}). 

The prefill phase refers to the initial processing of the prompt. During this step, all input tokens are encoded in parallel to compute hidden representations and initialize the key/value caches for attention. 

Decoding time refers to the average time the model requires to generate each new output token after the prefill phase, leveraging the cached representations of previous tokens.
\begin{table}[h!]
\centering
\resizebox{\columnwidth}{!}{%
\begin{tabular}{lcc}
\toprule
\textbf{Tokens kept} & \textbf{Prefilling time (ms)} & \textbf{Decoding time (ms/token)} \\
\midrule
100\% tokens & $162.54 \pm 3.07$ & $26.97 \pm 0.68$  \\
50\% tokens  & $34.37 \pm 0.57$  & $25.74 \pm 0.17$ \\
25\% tokens  & $29.55 \pm 0.20$  & $25.59 \pm 0.17$ \\
10\% tokens  & $26.89 \pm 0.15$  & $25.52 \pm 0.13$ \\
\bottomrule
\end{tabular}%
}
\caption{Average processing time per sample on Clotho v2 using Audio Flamingo 3 with a single A100 GPU. Results are reported as mean $\pm$ standard deviation.}
\label{tab:time}
\end{table}

%Naturally, we observe that the generation time per token remains approximately constant, independent of the size of the initial context. However, there is a significant reduction in the prefilling time, which becomes four times faster when the number of tokens is reduced by only a factor of two.
%jf: qqq corrections de forme

% PS2 : je retirerai obviously 
We observe that the generation time per token remains approximately constant, regardless of the size of the initial context. However, there is a significant reduction in time for the prefilling task, which becomes four times faster when the number of tokens is reduced by only a factor of two.
% PS : se démarquer de speechprune ici. 
% JF -> Petite conclusion ici ? A venir ? 
% PS2 : travail sur l'adapter par ex?

\section{Conclusion}

% PS2 :  by up to 4 times? pas fan du 2x-4x
% Rappeler le nom de la méthode que tu proposes. 
% Retirer "small" 

In this work, we showed that in audio–language models, pruning can reduce the audio encoder’s output by a factor of 2 to 4 while maintaining strong performance on standard evaluation metrics. We verified that this reduction of the number of audio tokens translates into a reduction of the prefilling time during inference of the audio-language model. We further introduced the Segmentwise Top-K method that better respects the signal’s sequential structure and yields consistent improvements on the benchmarks we studied.

As future work, we will explore variable-rate processing, adaptively selecting more or fewer tokens based on their estimated importance, and more explicitly leveraging temporal structure to further improve efficiency–accuracy trade-offs.

\vfill\pagebreak
\section{Acknowledgements}
This work was performed using HPC resources from GENCI IDRIS (Grant 2025-AD011014982R1). 
% References should be produced using the bibtex program from suitable
% BiBTeX files (here: refs, manuals). The IEEEbib.bst bibliography
% style file from IEEE produces unsorted bibliography list.
% -------------------------------------------------------------------------
\bibliographystyle{IEEEbib}
\bibliography{refs}

\end{document}